\documentclass{article}
\makeatletter

\@addtoreset{equation}{section}
\makeatother
\usepackage{epsf}
\begin{document}
\baselineskip 12pt
\title{The $f_0(980)$ and $a_0(980)$ Productions  \\
in Two-Photon Collisions and Radiative $\phi$ Meson Decays}

\author{M. Uehara\thanks{e-mail: ueharam@cc.saga-u.ac.jp}\\
Takagise-Nishi 2-10-17, Saga 840-0921, Japan}
\maketitle
\begin{abstract}
We study  how the scalar $f_0(980)$ and $a_0(980)$  states are produced 
in the two-photon collisions and radiative decays of the $\phi$ meson
through unitarized Born amplitudes with the charged pion and kaon loops 
followed by S-wave meson-meson scattering amplitudes. 
We found in a previous paper that $f_0(980)$ is generated as the bound 
state resonance, but $a_0(980)$ as the cusp, and find here that the nature 
of the generation of both states is consistent with the features of the 
production processes.
 \end{abstract}
\def\beq{\begin{equation}}　\def\eeq{\end{equation}}
\def\beqa{\begin{eqnarray}}　\def\eeqa{\end{eqnarray}}
\def\beqan{\begin{eqnarray*}}　\def\eeqan{\end{eqnarray*}}
\def\ba{\begin{array}}　\def\ea{\end{array}}
\def\noeq{\nonumber}
\def\mpi{m_\pi} \def\fp{f_\pi}　\def\mK{m_K} \def\fK{f_K}
\def\me{m_\eta} \def\fe{f_\eta}　\def\half{\frac{1}{2}}　
\def\der{\partial}
\def\vg{\mbox{\boldmath$g$}}　\def\vt{\mbox{\boldmath$t$}}
\def\vG{\mbox{\boldmath$G$}}　\def\vK{\mbox{\boldmath$K$}}
\def\vS{\mbox{\boldmath$S$}}   \def\vT{\mbox{\boldmath$T$}}
\def\vrho{\mbox{\boldmath$\rho$}}
\def\del{\delta}
\def\im{{\rm IM}} \def\re{{\rm R}}

\section{Introduction}
It has long been expected that the nature of the scalar $f_0(980)$ 
and $a_0(980)$ mesons is revealed by the investigation of the production 
processes such as the two-photon collisions and the radiative decays of 
the $\phi$ meson\cite{AchasovIvan,Kumano,Pennington,CloseTorn}. 
This is because these reactions offer valuable information which cannot 
be obtained in the meson-meson scattering processes alone.  
For example, the $f_0(980)$ resonance is  hidden 
in the broad $\pi\pi$ enhancement in the case of $\pi\pi$ scattering, but 
displays the peak in the two-pion invariant mass distributions 
in the cases of the radiative $\phi$ decay and the two-photon collisions.   
The magnitudes  of the two-photon partial widths, 
$\Gamma_{\gamma\gamma}(f_0)$ and $\Gamma_{\gamma\gamma}(a_0)$, 
and  the branching ratios, $B(\phi\to \gamma f_0)$ and 
$B(\phi\to\gamma a_0)$, are said to be useful 
to know the quark contents of the $f_0(980)$ and $a_0(980)$ states. 
The behavior of the amplitude 
$K^+K^-\to \pi^0\pi^0$ or $\to \pi^0\eta$ in the off-shell energy region 
below $K\bar K$ threshold could be seen through these reactions. 

So, we  study the two-meson mass distributions  by using the low energy 
S-wave scattering amplitudes obtained  in a previous paper\cite{Uehara}, 
which is referred to as I hereafter, where we showed that the amplitudes  
reproduce well,  though not quantitatively, the  low 
energy S-wave scattering behaviors. 
In order to give the scattering amplitudes we adopted the 
Oller-Oset-Pel\'aez version\cite{OOP} of the inverse amplitude method 
applied to the chiral perturbation theory\cite{NP}. 
We stressed in I that the $f_0(980)$ and $a_0(980)$ states 
are not twins, though the masses are very close to each other: 
The $f_0(980)$ state is generated as a typical bound state resonance 
grown from a bound state born in the $K\bar K$ channel, 
while the $a_0(980)$ state is not the bound state resonance 
but a typical cusp generated by the channel coupling between the 
$K\bar K$ and $\pi\eta$ channels,  where since the channel 
coupling is not sufficiently strong,  the cusp cannot develop into 
the resonance.  
It is interesting, therefore,  to study how the two states are 
produced from the same initial $\phi$ meson  and two-photon state, 
and  to see the structure of the off-shell $K\bar K\to \pi\pi$ and 
$K\bar K \to \pi\eta$ amplitudes. \\
 
Through the study of the  production processes we find the following 
results; 
\begin{enumerate}
\item  The clear $f_0(980)$ peak structure near $K\bar K$ threshold  
is seen in  the $\pi^0\pi^0$ mass distribution of both processes, but 
its width is too narrow in our model, reflecting the too 
steeply rising  behavior of the phase shift of the isoscalar $\pi\pi$ 
scattering amplitude in I.
\item We obtain $\Gamma_{\gamma\gamma}(f_0)B(f_0\to\pi\pi)=0.45$ 
keV by a rough estimate, where $B(f_0\to\pi\pi)$ is the branching ratio 
of the $f_0$ decay to the isoscalar $\pi\pi$ channel, and 
$B(\phi\to\gamma\pi^0\pi^0)=0.60\times 10^{-4}$, where the dominant 
contribution comes from the $f_0$ peak.
\item The $\pi\eta$ mass spectrum shows a clear cusp-like peak in 
the two-photon collisions, but a round one with a precipice at the 
$K\bar K$ threshold in the $\phi$ meson decay. Thus the shape of the 
$\pi\eta$ mass distribution changes drastically depending on the reactions. 
\item We obtain $\Gamma_{\gamma\gamma}(a_0)B(a_0\to\pi\eta)=0.35$
keV, and  $B(\phi\to\gamma\pi^0\eta)=1.15\times 10^{-4}$. 
\item  The $f_0(980)$ peak comes from  the amplitudes 
$K^+K^-\to \pi^0\pi^0$ and $K^+K^-\to \pi^0\eta$ below the $K\bar K$ 
threshold. In spite of the too narrow $f_0$ shape, we get naturally the 
ratio $B(\phi\to\gamma\pi\pi)/B(\phi\to\gamma\pi^0\eta)\,>\,1$.
\item The $ K^0\bar K^0$ peaks just above the $K\bar K$ threshold in 
both two-photon and $\phi$ decay processes come dominantly 
from the isoscalar $K\bar K$ elastic amplitude, and we get  
 $B(\phi\to\gamma K^0\bar K^0)=4.21\times 10^{-7}$. 
\end{enumerate}

In this paper we study the productions by the two-photon collisions in 
the next section, and by the radiative $\phi$ meson decays in Sec. 3, and 
finally we remark the relations of our findings to the experimental 
analyses and other theoretical works. 

\section{Two-meson production in the two-photon collisions}
Many theoretical papers have already been published on the 
$\gamma\gamma\to M\bar M$  reactions, of which  we cite a few 
recent ones; the elaborate amplitude analysis of 
$\gamma\gamma\to \pi\pi$ reactions by Boglione and 
Pennington\cite{Pennington} 
and the calculations on various final $M\bar M$ states using the 
amplitudes obtained by  the Bethe-Salpeter equation with chiral loops 
by Oller and Oset\cite{OO}, and references therein.  
In the above  $M\bar M$ denotes $\pi^+\pi^-$, $\pi^0\pi^0$, $\pi^0\eta$, 
$K^+K^-$ and $K^0{\bar K}^0$ final meson states. 

In order to understand how the similarity or 
dissimilarity of the $f_0(980)$ and $a_0(980)$ states appears in the 
relevant reactions, we adopt the formalism developed by 
Mennessier\cite{Mennessier}, where only the charged pion and kaon loops 
are taken into account.  
According to Mennessier the production amplitudes are given as the Born 
term + unitarized Born amplitudes. The production amplitude $F_f(w)$ 
from the initial photon state with the helicity $(++)$ to the the final 
$(M_1M_2)_f$ state is written as
\beq
F_f(w)=2e^2\left[B_f(w)+\sum_{i=1,2}T_{fi}(w)
G_i(w),\right] \label{ProdAmp}
\eeq
where $w$ is the total CM energy, $e$ the electric charge put outside 
the square brackets, and $i=1$ for $\pi^+\pi^-$ and $i=2$ for $K^+K^-$ 
channel.  The term, $B_f(w)$, is the S-wave Born term, which is followed 
only by the final $f=\pi^+\pi^-$ and $K^+\bar K^-$ 
states, and written as
\beqa
B_f(w)&=&\frac{1-\sigma^2_f(s)}{2\sigma_f(w)}
\log\left(\frac{1+\sigma_f(w)}{1-\sigma_f(w)}\right), \label{Born}\\
\sigma_f(w)&=&\sqrt{1-\frac{4m_f^2}{w^2}}\mbox{   for  }
f=\pi\pi\mbox{ and }K\bar K 
\eeqa
We normalize $T_{ij}$ 
with the  definite isospin so as to satisfy the following 
unitarity relation;
\beqa
{\rm Im}T^{(I)}_{ij}&=&-{T^{(I)}_{ik}}^*\rho_k(w)T^{(I)}_{kj}, \\
\rho_k(w)&=&\frac{\sigma_k(w)}{16\pi}\theta(w-w_k)
\eeqa
with $w_k$ being the threshold of the $k-$th channel.\footnote{ For $k=\pi\eta$,
$\sigma_k=\sqrt{1-2(m_\eta^2+m_\pi^2)/w^2+(m_\eta^2-m_\pi^2)^2/w^4}.$
} 

$G_i(w)$ is the integral of the charged pion or kaon loop connecting 
the $\gamma\gamma$ state with the helicity 0 to the S-wave 
$(M_1M_2)_f$ state, which is given as 
\beq
G_i(w)=\frac{1}{(4\pi)^2}\left\{1+\frac{m_i^2}{w^2}\log^2\left(
\frac{\sigma_i(w)+1}{\sigma_i(w)-1}\right)\right\}. \label{triangle}
\eeq
 
The production amplitude $F_f$ satisfies the final interaction theorem,  
\beq
 {\rm Im}F_f(w)=-T_{fi}(w)\rho_i(w) F^*_i(w) ,\label{FSI}
\eeq
since ${\rm Im}G_i(w)=-\rho_i(w)B_i(w)$ is hold. 
The S-wave two-meson production cross section in 
the $f-$final  state is written as 
\beq
 \sigma^f_S(w)=2\pi\alpha^2\frac{2k_f}{w^3}
 \left|\tilde F_f(w)\right|^2
\eeq
with $F_f=2e^2\tilde F_f$, and $k_f$ is the CM momentum. 
 
We notice that this production amplitudes do not involve any   
parameters which must be determined phenomenologically. 
Of course we could use generalized Born terms with vector meson 
exchanges between two photon vertices, but we ignore them. 
Fortunately, the contributions from the generalized Born terms are 
shown to be not significant below 1 GeV\cite{Mennessier}, where our 
interest is focussed on. 
\begin{figure}[h!]
\begin{center}
 \epsfysize=5cm
 \centerline{\epsfbox{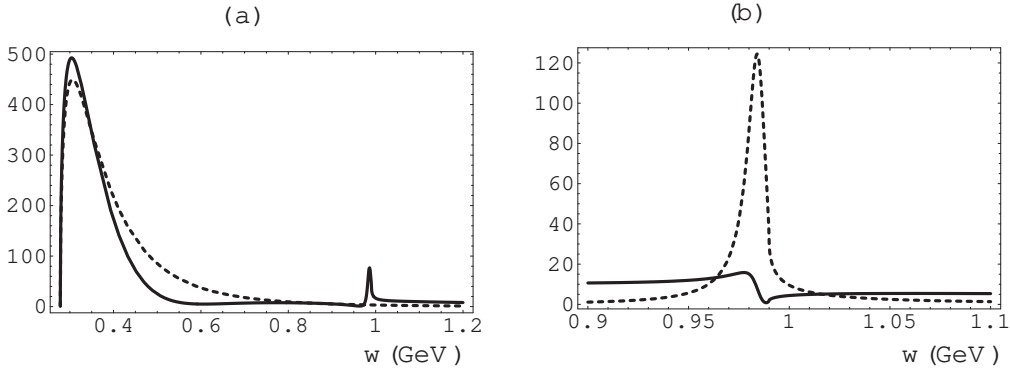}}
 \label{fig:chargedpipi}
 \caption{(a) 
 The cross section of the final $\pi^+\pi^-$ state in units of nb, and  
 the dotted line is cross section by the Born term. 
 (b) The contributions from $\pi^+\pi^-\to \pi^+\pi^-$ amplitude (solid line) 
 and the $K^+K^-\to\pi^+\pi^-$ amplitude(dotted line).}
\end{center}
\end{figure}
\begin{figure}[h!]
\begin{center}
\epsfysize=5cm
\centerline{\epsfbox{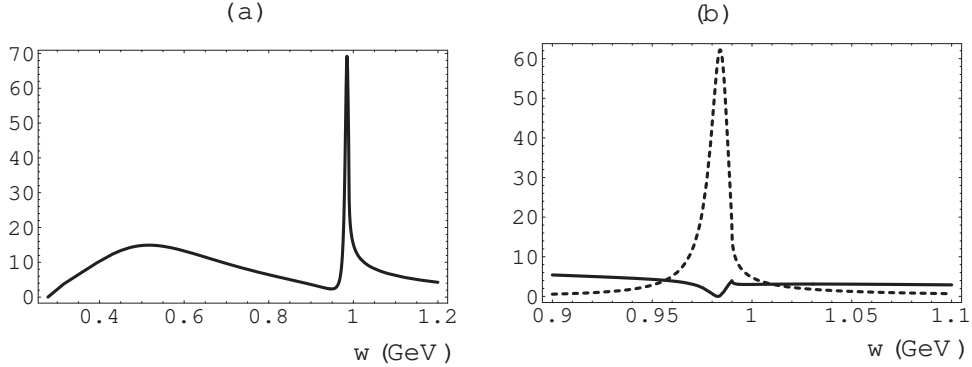}}
\label{fig:neutralpipi}
 \caption{(a) The  cross section of the final $\pi^0\pi^0$ state in units of nb.
 (b) The contributions from $\pi^+\pi^-\to \pi^0\pi^0$ amplitude (solid line) 
 and the $K^+K^-\to\pi^0\pi^0$ amplitude(dotted line). Note that the 
 height of the latter contribution is just a half of the counterpart of 
 the charged pion pair reflecting the isospin factor.}
 \end{center}
 \end{figure}
 \begin{figure}[h!]
 \begin{center}
 \epsfysize=5cm
 \centerline{\epsfbox{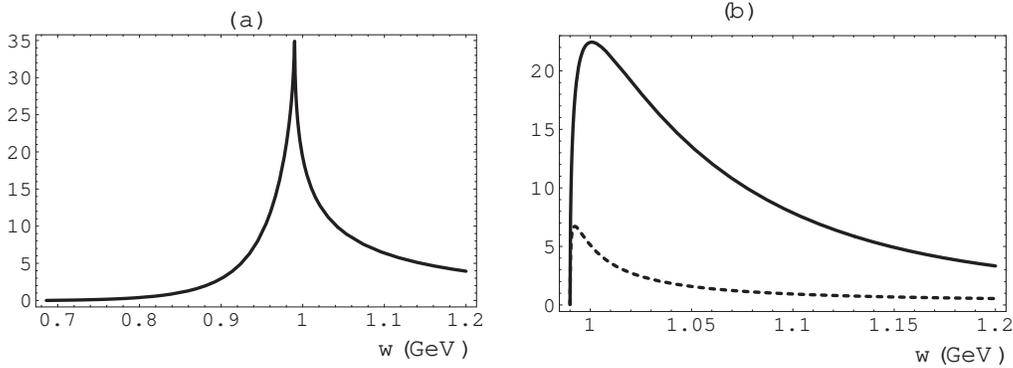}}
 \label{fig:pieta}
 \caption{(a) The  cross section of the final $\pi^0\eta$ state in units of nb.
 (b) The cross section for $K^+K^-$ state (solid line) 
 and the $K^0\bar K^0$ state(dotted line). }
 \end{center}
 \end{figure}
 The calculated results on the mass distributions of the two meson final 
states are shown in Fig. 1 to 3.  Our results are summarized as follows;
\begin{enumerate}
\item The $f_0(980)$  peak is seen in both  $\pi^+\pi^-$ and 
$\pi^0\pi^0$ spectra. This peak comes from the  off-shell amplitude 
$K^+K^-\to \pi\pi$ as shown in Fig.1b and 2b, while  
the on-shell amplitude $\pi^+\pi^-\to \pi\pi$ shows a dip similar to 
the $\pi\pi$ scattering cross section. The size of the off-shell peak of 
the charged state is as twice as that of the neutral one, 
respecting the isospin factor, but the large  interference between 
the pion and kaon loop-amplitudes makes 
the charged peak reduced to almost the same size as the neutral one. 
The shape of the peak is too narrow in our model. 
\item The small bump centered at 500 MeV in the $\pi^0\pi^0$ mass 
distribution reflects straight that of the isoscalar 
$\pi\pi$ cross section, for there are no contributions from the Born term 
and  the $K^+K^-\to\pi^0\pi^0$ amplitude almost vanishes there. 
\item The cusp-type peak of the $a_0(980)$ state appears in the $\pi^0\eta$ 
final state. The height of the peak is about a half of the $f_0(980)$ 
one in the $\pi^0\pi^0$ state. 
\item  The threshold peaks appear in both of the charged and neutral 
$K\bar K$ states; the large charged peak is due to the Born term, while the 
neutral one is due to the behavior of the isoscalar and isovector 
$K\bar K$ amplitudes near the threshold, where the isoscalar amplitude 
dominates over the isovector one.  
\end{enumerate}

After Ref.\cite{Pennington} we estimate roughly 
$\Gamma_{\gamma\gamma}(R)$ of the resonance  $R=f_0(980)$ and 
$a_0(980)$ using the formula 
\beq
\Gamma_{\gamma\gamma}(R)=
\frac{\sigma(\gamma\gamma\to f)_{\rm peak}m_R^2
\Gamma_{\rm tot}}{8\pi B_f},
\eeq
where $\sigma(\gamma\gamma\to f)_{\rm peak}$ is the cross section at  
$m_R$ of the $R$ resonance, $\Gamma_{\rm tot}$ the total width of $R$, 
and $B_f$ the branching ratio of $R$ to the final state $f$. 

As to $a_0(980)$ we take values, $m_{a_0}=0.990$ GeV,  
$\Gamma_{a_0}=0.1$ GeV, though the $a_0$ is not well expressed by 
the Breit-Wigner formula, and $\sigma_{\mbox{at peak} }=35$ nb. 
These value gives 
$\Gamma_{\gamma\gamma}(a_0)B(a_0\to\pi\eta)=0.35$ keV. 
It is difficult to estimate he branching ratio of the $a_0$. cusp to 
the $\pi\eta$ channel, so that $0.35$ keV is the lower bound. 

It seems rather difficult to estimate $f_0(980)\to \gamma\gamma$ 
partial width from the charged pion pair because of the large interference 
between the signal and the background 
as stated above, and then we use the cross section of the neutral pion 
pair, where the interference is small.  
Taking values $m_{f_0}=0.985$ GeV, $\Gamma_{f_0}=0.03$ GeV, 
$B_f=1/3\cdot B(f_0\to\pi\pi)$ and $\sigma_{\mbox{at peak} }=70$ nb, we have 
$\Gamma_{\gamma\gamma}(f_0)B(f_0\to\pi\pi)=0.63$ keV.  
Instead, if we add the charged and neutral peaks, that is 
$\sigma_{\mbox{at peak}}=150$ nb, 
we get $\Gamma_{\gamma\gamma}(f_0)B(f_0\to\pi\pi)=0.45$ keV.

While we should notice that our values are tentative and should not be taken 
seriously, because we do not attempt to get the best fits of the data,   
we observe that the partial widths  of $f_0(980)$ and
 $a_0(980)$ to the two-photon state are  almost equal   
 or the $f_0(980)$ partial width be larger than the $a_0(980)$ one 
 within our model.  
The heights of the $f_0$ peaks in the charged and neutral pion pairs do not 
satisfy the isospin multiplicative relation  $2\,:\,1$ because of  the 
interference between the elastic $\pi\pi$ amplitudes including $I=2$ 
component.

\section{The radiative decay of the $\phi$ meson}
Many theoretical works are based on the common production mechanism,   
in which the final S-wave two-meson state is produced following the 
charged kaon loop emitting a photon under the OZI rule in  the  
$\phi$ meson decay. This is the same 
mechanism as the one of the two-photon collisions, since it reduces to the 
radiative decay of the $\phi$ meson if one of the 
initial photon becomes massive and the other an outgoing photon.  
The validity of the kaon-loop mechanism is discussed by 
Achasov\cite{AchKloop}. 
The difference of the model depends on that of the 
 amplitudes used, for example  a unitary chiral 
approach\cite{Hirenzaki},  the Linear sigma 
Model\cite{Bramon},  phenomenological approaches 
with the Breit-Wigner formula\cite{Kumano,AchasovGubin,Golap} 
and references therein. 
We use the S-wave amplitudes $T(K^+K^-\to M_1M_2)$ obtained in I as 
the same as  in the preceding section. 
The interference with the $\phi\to\rho\pi^0\to \gamma\pi^0\pi^0$ 
sequential decay amplitude is ignored in this paper, though 
the experimental data do not exclude the contamination of the sequential 
processes. This is discussed in Refs.\cite{AchasovGubin, Bramon}. 

According to Ref.\cite{Hirenzaki} we write the decay amplitude of the 
$\phi$ meson to two-meson state $\gamma M_1M_2$, where $M_1M_2$ 
is $\pi^0\pi^0$, $\pi^0\eta$ and $K^0\bar K^0$ as
\beq
F(\phi\to\gamma M_1M_2)=
2eg \{ G_K(w)+f\frac{m_\phi^2-w^2}{2m_\phi^2}g_K(w) \}
T_{K^+K^-\to M_1.M_2}, \label{decayamp}
\eeq
where  $g$ and $f$ are the parameters  defined as 
\beq
g= \frac{G_Vm_\phi}{\sqrt{2}f_\pi^2}\quad\mbox{ and }\quad 
f=\frac{F_V}{2G_V}-1
\eeq
with $G_V$ and $F_V$ being the constants in the chiral 
Lagrangian\cite{Ecker,Huber,Hirenzaki} and a factor $1/\sqrt{2}$ in $g$ 
is the ratio of the coupling constant $\rho\pi\pi$ to $\phi K\bar K$. 
We adopt the values $G_V=55$ MeV and $F_V=165$ MeV given in 
Ref.\cite{Hirenzaki}, which give $g=4.69$ and $f=0.5$.
The triangle $K\bar K$  loop integral $G_K(W)$, which connects the 
$\phi$ meson to {\em the S-wave two-meson state} after emitting a 
photon, is given as 
\beqa
G_K(w)&=&\frac{1}{(4\pi)^2}\left\{1+\frac{m_k^2}{w^2-m_\phi^2}
\left[\log^2\left(\frac{\sigma_K(w)+1}{\sigma_K(w)-1}\right)-
\log^2\left(\frac{\sigma_K(m_\phi)+1}{\sigma_K(m_\phi)-1}
\right)\right] \right. \\ \noeq
&&\left.-\frac{m_\pi^2}{w^2-m_\phi^2}\left[\sigma_K(w)
\log\left(\frac{\sigma_K(w)+1}{\sigma_K(w)-1}\right)
-\sigma_K(m_\phi)\log\left(\frac{\sigma_K(m_\phi)+1}
{\sigma_K(m_\phi)-1}\right)\right] \right\},
\eeqa
and the loop integral $g_K(w)$ by
\beq
g_K(w)=\frac{1}{(4\pi)^2}\left\{-1+\log\left(\frac{m_K^2}{\mu^2}\right)+
\sigma_K(w)
\log\left(\frac{\sigma_K(w)+1}{\sigma_K(w)-1}\right)\right\},
\eeq
where we take $\mu=1$ GeV as in I. 
One should note that the above triangle loop integral is the same as 
$I(a,b)(a-b)/8\pi^2$ with $I(a,b)$ being defined in 
Ref.\cite{AchasovIvan,Bramon1,Kumano}, and 
that as $m_\phi\to 0$ this reduces to the triangle loop integral $G_i(w)$ 
given in Eq.(\ref{triangle}) for $\gamma\gamma \to MM$ used in the previous 
section.

The mass dependence of the two-meson state in the radiative decay 
is given as 
\beq
\frac{d\Gamma(w)}{dw}=\left(\frac{\alpha}{3\pi}\right)
\left(\frac{g^2}{4\pi}\right)\frac{k_f(m_\phi^2-w^2)}{m_\phi^3}
\left|\tilde F(\phi\to\gamma M_1M_2)\right|^2,
\eeq
where $k_f$ is the momentum of the final $(M_1M_2)_f$ state in the 
two-meson CMS, and we define $\tilde F$ as $F=2eg\tilde F$. The $w$ 
dependence of the phase space factor $k_f(m_\phi^2-w^2)/m_\phi^3$ has 
the maximum  near the middle of the whole $w-$range, and 
vanishes at both of the ends. Furthermore, the loop integral part 
$G_K(w)+f(m_\phi^2-w^2)/(2m_\phi^2)\cdot g_K(w)$ has a strong cusp 
behavior  at the $K\bar K$ threshold, which is very near to $m_\phi$, 
so that the mass distributions near 1 GeV are affected double by 
these kinematical factors. The term with $f$ in the loop integral 
plays a role to reduce the first $G_K$ term below the $K\bar K$ threshold 
for $f\,>\,0$. 
\begin{figure}[h!]
\begin{center}
\epsfysize=5cm
\centerline{\epsfbox{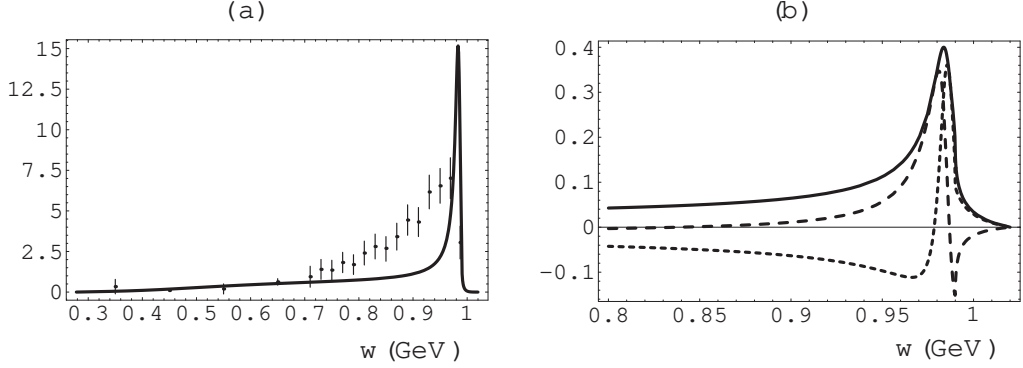}}
\label{fig:phipipi}
 \caption{(a) The $\pi^0\pi^0$ invariant mass distribution normalized 
 by $\phi$ meson total width, $dB(\gamma\pi^0\pi^0)/dw\times 10^7$/MeV. 
Experimental data are taken from SND data\cite{SNDpipi}. 
(b) $F(\gamma\phi\to \pi^0\pi^0)$ amplitude; the real part is given 
 by the dashed, the imaginary part by the dotted and the absolute value 
 by the real line.}
 \end{center}
 \end{figure}
\begin{figure}[h!]
\begin{center}
\epsfysize=5cm
\centerline{\epsfbox{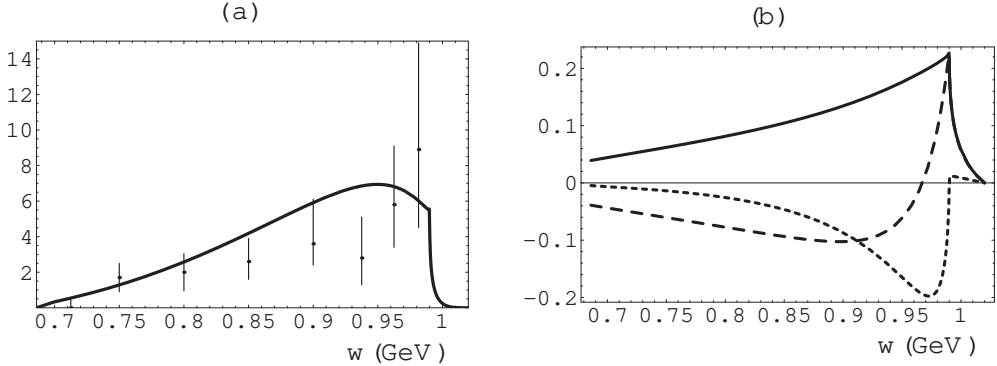}}
\label{fig:phietapi}
 \caption{(a) $dB(\gamma\pi^0\eta)/dw\times 10^7$/MeV. 
 Experimental data are taken from SND data\cite{SNDpieta}
 (b) $F(\phi\to \gamma\pi^0\eta)$ amplitude; the real part is given 
 by the dashed, the imaginary part by the dotted and the absolute 
 value by the real line.}
 \end{center}
 \end{figure}
\begin{figure}[h!]
\begin{center}
\epsfysize=5cm
\centerline{\epsfbox{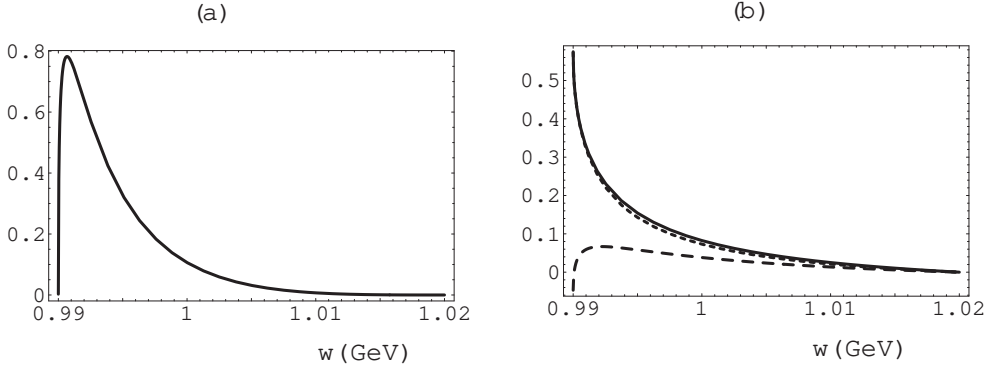}}
\label{fig:phiK0K0}
 \caption{(a) $dB(\gamma K^0\bar K^0)/dw\times 10^7$/MeV. 
 (b) Lines are the same as in previous Figs. }
 \end{center}
 \end{figure}
The calculated mass distributions and production amplitudes 
are plotted in Figs. 4 to 6, in which we find the following results:
\begin{enumerate}
\item The $f_0(980)$ peak is manifest in the final 
$\pi^0\pi^0$ mass distribution as shown in Fig.4. The integrated 
branching ratio, $B(\phi\to\gamma\pi^0\pi^0)=0.60\times 10^{-4}$, 
and then we have $B(\phi\to \gamma\pi\pi)=1.80\times 10^{-4}$, where 
the dominant contribution comes from the $f_0$ peak.   
\item Contrary to the $f_0(980)$ peak, the $\pi^0\eta$ mass distribution 
does not show the cusp like peak but the round peak accompanied with 
the precipice at the $K\bar K$ threshold . This is due to
the kinematical factors pointed out in the above. 
The integrated branching ratio is $B(\gamma\pi^0\eta)=1.15\times 10^{-4}$. 
\item  The $K^+K^-\to K^0\bar K^0$ amplitude is dominated by 
the isoscalar $K\bar K$ elastic amplitudes and is not small compared 
with the amplitudes at the $K\bar K$ threshold as shown in Fig. (6b). 
We get $B(\phi\to \gamma K^0\bar K^0)=4.21\times 10^{-7}$.
 \end{enumerate}

 \section{Concluding Remarks}
 Here we remark some relations of our results with the experimental 
 analyses and other theoretical works. 
 
\subsection{Two-photon collision processes }
Some experimental groups measured the two-photon partial width of the 
$f_0$ state $\Gamma_{\gamma\gamma}(f_0)$ through the 
fits of the $\pi\pi$ mass spectra by the Breit-Wigner form,
\beq
\sigma_{\gamma\gamma\to f_0\to \pi\pi}(w)=
8\pi\left(\frac{m_R}{w}\right)^3\frac{\Gamma_{\gamma\gamma}(f_0)
B(f_0\to \pi\pi)\Gamma_{\rm tot}(w)}
{(w^2-m_R^2)^2+m_R^2\Gamma_{\rm tot}^2(w)},
\eeq
where $m_R$ is the $f_0$ mass,  $\Gamma_{\rm tot}$ is the $f_0$ 
total width  and $B(f_0\to \pi\pi)$ the branching ratio into $\pi^0\pi^0$ 
or $\pi^+\pi^-$. Using $m=975$ MeV, $\Gamma_{\rm tot}(m)=33$ MeV and 
$B(f_0\to\pi^0\pi^0)=1/3\times 0.78$, 
the Crystal Ball Collaboration gives 
$\Gamma_{\gamma\gamma}(f_0)= 0.31\pm 0.17$ keV\cite{Crystalpipi}.  
JADE Collaboration reports  $\Gamma_{\gamma\gamma}(f_0)\cong 0.42$ 
keV from the $\pi^0\pi^0$ final state, though the significance of the 
$f_0(980)$ signal is not high\cite{JADE}. 
 From the $\pi^+\pi^-$ final state Mark II detector group gives  
 $\Gamma_{\gamma\gamma}(f_0)= 0.29\pm 0.11$ keV, where $m=1012$ 
 MeV,  $\Gamma_{\rm tot}(m)=52$ MeV and 
 $B(f_0\to \pi^+\pi^-)=2/3\times 0.78$ 
are used\cite{Boyer}. If we use the same mass and width 
of the $f_0$ state as the Crystal Ball Collaboration, we might have 
$0.17$ keV instead of $0.29$ keV, because 
$\Gamma_{\gamma\gamma}(f_0)$ is 
proportional to $m_R^2\Gamma_{\rm tot}$. The values $0.31$ keV in the 
$\pi^0\pi^0$ spectrum and $0.17$ keV in the $\pi^+\pi^-$ one may indicate  
that the ratio of the peaks of the charged and neutral pion pairs could  
not be simply 2 : 1 as we encountered in Sec.2. 

Although there is seen only a subtle signal of the $f_0(980)$ peak in the 
experimental two-pion mass spectra masked by the large tail of the 
$f_2(1279)$ resonance, the theoretical works including us predict 
the existence of the clear $f_0(980)$ peak over the $f_2$ 
background\cite{Pennington,OO}.  For example, the $f_0(980)$ peak 
in the peak solution of the $I=0$ cross section is about 180 nb as seen 
in Fig.17 of Ref.\cite{Pennington}. 
If the pion-loop dominates the whole energy region below 1 GeV, the 
$f_0(980)$ peak does not appear reflecting the structure of the 
$\pi\pi$ scattering amplitude, but we have to include 
the kaon-loop into the game. The kaon-loop is followed  by 
the $K\bar K\to\pi\pi$ transition amplitude, which displays a peak 
related to the $f_0(980)$ state {\em below the $K\bar K$ threshold} 
as shown in Figs.1b and 2b. Indeed such a peak 
is observed in the radiative $\phi$ decay into $\gamma\pi\pi$. The shape 
by our calculation is too narrow, but we hardly avoid the peak structure, 
because  the $f_0(980)$ state strongly  couple to the
$K\bar K$ channel. Theoretical estimate of the partial width of $f_0$ gives 
$0.13\sim 0.36$ keV in \cite{Pennington} and $0.20$ keV in \cite{OO}. 
But if we make the same rough estimate for $\sigma_{\rm peak}=180$ nb, 
we have $\Gamma_{\gamma\gamma}(f_0)B(f_0\to\pi\pi)\cong 0.54$ keV.

The two-pion mass spectrum produced by two-photon collisions is a 
useful field to investigate the off-shell $K\bar K\to \pi\pi$ 
scattering amplitude and the production mechanism. 
We expect, therefore, that the partial wave analysis of the two-pion 
mass spectrum  in the region of 1 GeV is performed. \\
 
As to the $a_0(980)$ state the experimental data  by Crystal Ball  
and JADE Collaborations give a clear peak in the 
$\pi\eta$ spectrum\cite{Crystalpieta,JADE}. 
The peak of the cross section is about 40 nb, which is very similar to ours. 
The two-photon partial width is 
$\Gamma_{\gamma\gamma}(a_0)B(a_0\to\pi\eta)\cong 0.19$ 
keV\cite{Crystalpieta} and 
$\cong 0.28$ keV\cite{JADE}, which are a little smaller than our rough 
estimate, $0.35$ keV.  
They include a  non-resonant smooth background under the the $a_0$ 
peak, however, while our calculation uses the full cross 
section without any background. 
Further,  the $a_0(980)$ state could be expressed not by the 
Breit-Wigner form  but by a cusp, whose peak is sit just at the 
$K\bar K$ threshold.  
These may give an effect on the values of 
$\Gamma_{\gamma\gamma}(a_0)$. \\

The threshold peak of the 
$\gamma\gamma\to K^0\bar K^0$ cross section is about 7 nb, that is not 
so small compared with the $\pi\pi$ and 
$\pi\eta$ states. There are experimental data for $K_s^0K_s^0$ final 
state, which indicate about 2 nb at 1.1 GeV with a large error\cite{CELLO}, 
with which our prediction could  be consistent.

 \subsection{The radiative decays of the $\phi$ meson}
As to the final 
$\gamma\pi^0\pi^0$ state, all the experimental results have a peak 
near 970 MeV and a small tail below 600 MeV in the 
invariant $\pi^0\pi^0$ mass distributions. The branching ratio 
$B(\phi\to\gamma\pi^0\pi^0)$ over the whole mass range is  given  
$(1.158\pm 0.093\pm 0.052)\times 10^{-4}$ by the SND 
detector group\cite{SNDpipi}, $(1.08\pm 0.19\pm0.09)\times 10^{-4}$ 
by the CMD-2 group\cite{CMD2} and $(1.09\pm 0.03\pm 0.05)\times 10^{-4}$ 
by KLOE collaboration.  These values are consistent with each other. 
The theoretical work\cite{Hirenzaki} related to ours gives 
$B(\phi\to\gamma\pi^0\pi^0)=0.8\times 10^{-4}$, which is a little 
larger than, but not inconsistent with our calculation. 

The results become strongly model-dependent, however, when we  
extract the $B(\phi\to\gamma f_0(980))$ from the 
$\phi\to\gamma\pi^0\pi^0$ mass distribution. The SND group assumes 
that the whole mass distribution is dominantly given by the $f_0(980)$ 
resonance with $m_{f_0}=969.8\pm 4.5$ MeV and 
$\Gamma_{f_0\to\pi\pi}\approx 200$ MeV, and then gives 
\beq
B(\phi\to \gamma f_0)=(3.5\pm 0.3{}^{+1.3}_{-0.5})\times 10^{-4}. \label{SND}
\eeq
The CMD-2 group  gives
\beq
B(\phi\to\gamma f_0(980))=(3.05\pm 0.25\pm 0.72)\times 10^{-4}
\eeq
under the single $f_0(980)$ fit, but 
\beq
B(\phi\to\gamma f_0(980))=(1.5\pm 0.5)\times 10^{-4}
\eeq
under the two-resonance fit with $f_0(980)$ and $f_0(1200)$, which is 
smaller by a factor of 2 than the one under the single resonance fit. 
On the other hand KLOE collaboration fits the $\pi^0\pi^0$ mass 
distribution in terms of the $\gamma(f_0+\sigma)$ mode, and gives 
\beq
B(\phi\to \gamma f_0)=3\times(1.49\pm 0.07)\times 10^{-4}
\cong 4.47\times 10^{-4}, 
\eeq
where the group uses  $m_\sigma= 478$ MeV and 
$\Gamma_\sigma= 324$ MeV, which are set in Ref.\cite{Aitala}. 
The values of $B(\phi\to\gamma f_0)$ are, thus, scattered depending 
on the assumption adopted by each group, though the invariant 
mass distributions  coincide with each other. 

The last result by KLOE collaboration is much larger than others. 
According to their analysis the $\sigma$ 
contribution gives a bump near 500 MeV, but the experimental mass 
distribution is very small there, so that they are forced to erase the bump 
by the destructive interference with the $f_0$ resonance with a 
large width, where $m_{f_0}=973$ MeV and 
$\Gamma(f_0\to\pi\pi)=260$ MeV. 
This enlarges the $\gamma f_0$ branching ratio. 
On the other hand, Achasov and Gubin\cite{AchasovGubin} 
analyze the Novosibirsk data\cite{SNDpipi,CMD2} by assuming the 
$f_0(980)$ and $f_0(1500)$ resonances with the background phase 
$\delta_B(w)=b\sigma_1(w)$ with $b$ being a positive constant, 
which describe the $I=0$ $\pi\pi$ scattering phase shift rather well, 
and give $B(\phi\to\gamma f_0)=3\times(0.78\sim 1.01)\times 10^{-4}
\cong (2.34\sim 3.03)\times 10^{-4}$. 
Since the off-shell amplitude $T(K^+K^-\to\pi^0\pi^0)$ must have the same 
phase as the pure isoscalar $T^{(0)}(\pi\pi\to\pi\pi)$ amplitude below the 
$K\bar K$ threshold by the unitarity, it should be verified that such a sum of 
the $\sigma$ and $f_0(980)$ states can really give the isoscalar phase shift 
of $\pi\pi$ elastic scattering. 

We note that the QCD sum rule 
give $B(\phi\to\gamma f_0)=(2.7\pm 1.1)\times 10^{-4}$\cite{Fazio} 
and $(3.5\times(1\pm 0.3))\times 10^{-4}$ \cite{Aliev}. 

It is not reasonable to expect the large contribution from the 
$f_0(980)$ to the low mass region such as 500 MeV from the elastic 
$\pi\pi$ scattering behavior, so that 
$B(\phi\to\gamma f_0)$ could be much reduced to
$\sim 1.7\times 10^{-4}$, if the $f_0(980)$ contribution is restricted 
to the mass region larger than 900 MeV in the data of SND 
group\cite{SNDpipi}.  
It is dangerous, therefore, to discuss the nature of the $f_0(980)$ 
and $a_0(980)$ using the too model-dependent values. 

As to the $\gamma\pi^0\eta$ final states, two Novosibirsk groups 
SND\cite{SNDpieta}, CMD-2\cite{CMD2} and KLOE 
collaboration\cite{KLOEpieta} give $B(\phi\to\gamma\pi^0\eta)$  
a consistent value $(0.80\sim 0.90)\times 10^{-4}$ within the errors. 
The shapes of the invariant $\pi^0\eta$ mass distribution seem to be 
different, though the number of events maybe  not enough. The shape of the 
KLOE collaboration does not show a clear peak near the $K\bar K$ 
threshold, that is very similar to ours and other theoretical 
works\cite{Bramon}. The integrated branching ratio in 
Ref.\cite{Hirenzaki} is $0.87\times 10^{-4}$, which is a little smaller 
than ours, but not inconsistent with ours. 

Since there are no clear resonant states other than the $a_0(980)$ state 
below 1 GeV the branching ratio for $\gamma\pi^0\eta$ could 
be assigned to the $a_0(980)$ state as a whole, irrespective of the 
assumed mass and width of the $a_0$ state. On the other hand our 
calculation on the integrated branching ratio $B(\phi\to\gamma\pi^0\pi^0)$ 
comes almost from the $f_0$ peak, so that $B(\phi\to \gamma f_0)\sim 
3B(\phi\to\gamma\pi^0\pi^0)=1.8\times 10^{-4}$. 
Thus, it is not unreasonable to say that 
$B(\phi\to\gamma f_0)\,>\, B(\phi\to \gamma a_0)\sim 1\times 10^{-4}$.

The decay into the $\gamma K^0\bar K^0$ state may reveal the structure of 
the $K\bar K$ elastic amplitudes just above the $K\bar K$ threshold. 
Our result of the branching ratio is $4.2\times 10^{-7}$, that is 10 
times as large value as the 
one predicted by the $f_0(980)\pm a_0(980)$ resonance model written 
in terms of the Breit-Wigner forms\cite{AchasovKK}.  \\

We have studied how the $f_0(980)$ and $a_0(980)$ are produced in the 
two-photon collisions and the radiative decays of the $\phi$ meson. The 
structure of the production amplitudes common to both of the processes is 
that the production  proceeds through the charged pion- and kaon-loop 
diagrams.  It is understood by the same mechanism that the 
two-pion spectrum in the $J/\psi\to\phi\pi\pi$ show a peak near  
$f_0(980)$ resonance as like as the $\phi\to\gamma\pi\pi$ decay  process, 
if we respect the OZI rule at the decay vertices and $\phi$ be a spectator. 
Similarly, the bump 
peaked at 500 MeV in  the two-pion mass spectrum in the 
$J/\psi\to\omega\pi\pi$ comes from the pion-loop followed by 
$\pi\pi$ elastic scattering amplitude.
Thus, these processes offer a 
unique field to investigate the off-shell $K\bar K\to \pi\pi$ and 
$\to \pi\eta$ amplitudes accompanied by the pion- and kaon-loop 
diagrams. 

In I we studied how these scalar states are 
generated through a unitarized chiral perturbation theory. There we 
pointed out that the generation mechanism may be different from 
each other, and the $f_0$ state is the typical bound state resonance, 
but the $a_0$  the cusp generated by the channel coupling. 
We find that this nature of the generation is consistent with 
the features of the production processes. 
We, thus, conclude within our scheme that we have naturally 
$1\mbox{ keV}\,>\,\Gamma_{\gamma\gamma}(f_0)\,>\,
\Gamma_{\gamma\gamma}(a_0)$  and 
$B(\phi\to\gamma f_0)\,>\, B(\phi\to\gamma a_0)\sim 10^{-4}$. 

The experimental $\pi^0\pi^0$ invariant mass spectrum in the two-photon 
collisions show a small but broad bump centered at 500 MeV, though 
this bump does not play any leading role in our production processes  
in contrast to the $f_0$ and $a_0$ states. The bump is naturally reproduced 
by our model with S-wave $\pi\pi$ scattering amplitudes following the 
charged pion loop. We regard the bump not as the resonant state but as 
the vacuum excitation owing to the spontaneous break down of chiral 
symmetry.  We point out that the $\sigma$ resonance with a mass about 
480 MeV and a large width would not be consistent with these 
production processes nor the scattering processes. 

We cannot find any reason to change our point of view of the low mass 
scalar mesons developed in I through the study of the photon related 
production processes.

\begin{flushleft}
{\bf Acknowledgment}
\end{flushleft}
The author thanks Department of Physics,  Saga University for the 
hospitality extended to him. 

\end{document}